# Hydroanalysis of Animal Lysozymes *c* and Human Defensins α

## J. C. Phillips

Dept. of Physics and Astronomy, Rutgers University, Piscataway, N. J., 08854

**Proteins appear to be the most dramatic natural example of self-organized criticality (SOC), a concept that explains many otherwise apparently unlikely phenomena. Protein functionality is dominated by long range hydro(phobic/philic) interactions which both drive protein compaction and mediate protein-protein interactions. In contrast to previous reductionist short range hydrophobicity scales, the holistic Moret-Zebende hydrophobicity scale represents a hydroanalytic tool that bioinformatically quantifies SOC in a way fully compatible with evolution. Hydroprofiling identifies chemical trends in the activities and substrate binding abilities of model enzymes and antibiotic animal lysozymes *c* and antibiotic human defensins, which have been the subject of tens of thousands of experimental studies. The analysis is simple and easily performed, and immediately yields insights not obtainable by traditional methods based on short-range real-space interactions, as described either by classical force fields (CFF) used in molecular dynamics simulations (MDS), or hydrophobicity scales based on transference energies from water to organic solvents.**

Lysozymes are a ubiquitous protein family (1) which contains hen egg-white (HEW) lysozyme, probably the most studied protein. Lysozymes function both as enzymes and as antibiotics. The enzyme function is much studied, and is regarded as archetypical. It was originally (1965) supposed to be mediated by short range ionic interactions; it has only recently (2001) been identified as also mediated by short range covalent and H-bond interactions (2). Antimicrobial (ABM or lytic) mechanisms are much more complex, and are still the subject of many studies (3). Although HEW lysozyme (129 residues) is much larger than defensins (~ 30 residues), it shares essential structural and AMB functions. Using the Moret-Zebende hydrophobicity scale Ψ (4) based on the long range power-law evolution of solvent-accessible surface areas with increasing segmental length, one can compare chemical trends in AMB activity of defensins, lysozymes, and related proteins. The results for protein functionality obtained with the holistic long range MZ scale are consistently superior to those obtained from short range reductionist scales based on transference energies of isolated amino acids from water to organic solvents (5-7); the differences are especially pronounced in functionally critical regions (8-10). The insights obtained can be used to engineer new proteins with potentially desirable AMB properties (11); such insights are expected on fundamental grounds, as the first hydration monolayer exhibits distinctive properties (12).

**Systematics of Wild Lysozymes**

Much of the source material used here comes from two review articles (13,14). Most discussions of enzyme functionality begin with the traditional lock and key mechanism which emphasizes short range interactions, a picture that has been refined by successive mutagenic studies of enzyme cores, often complexed with a simple substrate (2). Thus the basic structure of lysozyme *c* consists of two lobes or domains (right, α helices, left, β strands) surrounding an active site cleft (Fig. 1) which binds six sugar rings (A-F). Modeling led to the insight that the nearby basic acid pair $Asp^{52}$ and $Glu^{35}$ could exchange charge to stabilize the transition state (or unstable intermediates) in the reaction mechanism (2). This primary model has since been supported in many ways; here it is revisited to determine the secondary long range factors that determine quantitative chemical trends in wild lysozymes. (13) lists the sequences of 75 animal and insect *c* lysozymes, from HEW to human; detailed study of these, primarily supported by the known structure and the MZ hydrophobicity scale, and secondarily by other scales (such as less accurate and contextually limited helical propensity scales (16)) enables us to recognize long-range interactions that are important for protein engineering. Equally important are the short range interactions studied by mutations (14); in practice one would engineer a combination dictated by these and other factors.

The lysozyme chemical trends studied here concern AMB or lytic activity against gram-positive bacteria, metabolic activity against glycol chitin, and binding ability against three sugar rings (N-acetylglucosamine, $(NAG)_3$). These are shown in Table I, reproduced here for the reader's convenience from (14), for seven examples. These examples fall naturally into two groups, birds and placental mammals, and when these are arranged in order of lytic activity, the groups are simply separated. This is not the case for the other two properties, and it is clear that evolution has aimed mainly at strengthening lytic activity (3), which is more complex than metabolic activity. The very large binding ability of HEW lysozyme against multiple sugar rings is striking: it appears that the HEW sequence is nearly perfectly adapted to this task, while the turkey lysozyme sequence, only slightly changed from HEW, is much weaker.

Of the 130 lysozyme *c* residues, 23 are invariant among animals, including 8 Cys residues that form 4 disulfide bonds (13), as well as the basic acid pair $Asp^{52}$ and $Glu^{35}$. (13) identified a region around invariant Trp residues 108 and 111, as well as 28, as a "hydrophobic region A", and a long loop sequence 50-76 as a "hydrophilic domain B". On the MZ scale Trp is not strongly hydrophobic (it is merely a typically hydrophobic "core" residue), and the averages over X= Trp 28, 108 and 11 of MZ $<\Psi3>(X)$ are 0.167 (HEW) and 0.165 (human), in other words, little changed: while important for stability, Trp is not critical to lysozyme functionality (15). Domain B (the long loop sequence 50-76) is indeed hydrophilic in HEW ($<\Psi27>$ = 0.147) but in human lysozyme, $<\Psi27>$ = 0.153 (quite close to hydroneutral (0.155)), so this loop is possibly a factor in the large HEW sugar binding ability, which has weakened in human lysozyme.

**Wild HEW, Turkey and RN (Japanese) Pheasant Lysozymes**

We can test ideas like the importance of the hydrophilicity of Domain B by combining the MZ scale (with its long-range SOC accuracy) with the detailed changes in properties of these three birds, as listed in Table 1. As expected, there are few sequence changes between them, and only two single mutations in Domain B. Overall most of the mutations are singles, many of which are innocuous (for example, HEW R73 → (TUR and PHE)K73 does not alter charge or hydrophobicity significantly). There is virtually no change in <Ψ27> for Domain B between HEW, TUR, and PHE, which means that after all this loop is not an important factor in weakening the large HEW sugar binding ability (which decreased by approximately a factor of 9 between HEW and TUR, Table 1).

There are three significant HEW → TUR hydrophobicity changes: RH$^{15}$G → RL$^{15}$G, which changes <Ψ3>(15) from 0.129(HEW) to 0.144(TUR); TQ$^{41}$A → TH$^{41}$A, which changes <Ψ3>(41) from 0.132(HEW) to 0.148(TUR), and VQ$^{121}$A → VH$^{121}$A, which changes <Ψ3>(121) from 0.167(HEW) to 0.182(TUR). Individually these changes are all small (because only one site is mutated, and <Ψ3> is averaged over three sites), but they all have the same sign and are in fact nearly equal (the TUR sites are all more hydrophobic). These three sites lie well outside the sugar-binding cleft and the hydrophobic core, on the protein surface. The nearly perfect compaction of HEW (400 million years old (1)!) by hydrophobic forces suggests that the water monolayer of HEW is an exceptionally well adapted (stress-free) glassy network, similar to the nearly ideal networks observed in the reversibility windows of network glasses (17). Another interpretation of this near equality of hydrophobicity changes is that the interfacial water-protein surface tension is constant around the entire protein surface (as it has to be if the protein sequence is evolutionarily adapted to optimize these interactions, and as Gibbs would have hoped for his droplet model). These three hydrophobic mutations in TUR cooperatively disrupt the HEW water monolayer network and thereby stiffen the TUR backbone, so that its ability to bind multiple sugar rings is much weaker than HEW's. At the same time the lytic activity of TUR lysozyme is enhanced (Table 1), probably because this depends on lysozyme cooperativity.

The next case is HEW → PHE. Here there are seven significant mutations, and just as with TUR, they occur outside the hydrophobic core on the protein surface. All seven changes are small, of similar magnitudes, and the same sign, much as for HEW → TUR, implying strong support for the interfacial water-protein surface tension mechanism discussed above. Thus one would expect to see differences in properties of HEW and PHE to be twice as large as for TUR and HEW, but according to Table 1, they are only about 1/3 as large. What has happened? The answer is that these seven mutations are divided into two subsets, with four mutations distant from Cys disulfide bridges, and three mutations adjacent to bridges (denoted by *). Disulfide bridges cause otherwise singly connected protein chains to be multiply connected. Modified hydrophobic interactions near disulfide bridges can strengthen these bridges and increase the ability of PHE relative to TUR to bind sugar rings, thus bringing it closer to HEW in properties. In fact, one can define a mutational configuration coordinate M

$$< <\Psi 3>_M ** > = <<\Psi 3>_M> - 2 <<\Psi 3>_{M*} > \qquad (1)$$

with $<\Psi 3>$ averaged over mutations distant from disulfide bridges, and $<\Psi 3>*$ averaged over mutations adjacent to bridges. The factor of 2 in (1) is what one would expect if the hydrogen bonds involved in water monolayer-protein interfaces are stabilized quantum mechanically, and are out of phase at branching disulfide bridges, or if internal stresses are balanced at the multiply connected bridging sites. Sugar binding correlates well with $<\Psi 3>_M **$.

The discussion leading to Eqn. (1) is quite abstract, and it may seem unjustified to many readers; indeed, like much abstract mathematics, it requires more time to understand than to read. The key points are the nearly equal hydrophobic steps between HEW and PHE that occur for different wild sequences at seven spatially distant, superficially unrelated sites on an essentially common lysozyme surface: this is the hydrophobic analogue of the classical concept of protein water interfacial surface tension. (This concept assumes that the tension is nearly constant over the surface, which is necessary if this interaction makes the dominant contribution to stabilizing the main features of the surface geometry.) As in Gibbs nucleation models of first-order phase transitions, surface or interfacial tension is expected to be an essential factor determining functional properties of proteins regarded as self-organized networks near critical points. Such equality is of course very unlikely *a priori*, but one can go further. The MZ scale itself is based on exponents from power-law fits to the long range length dependence of solvent-accessible surface areas (4), so that its underlying justification is SOC: this is the unifying holistic mechanism that leads to nearly equal hydrophobic steps for spatially distant sites on an essentially common lysozyme surface.

To test this idea, one can repeat the calculations using one of the many reductionist hydrophobicity scales based on transference energies of individual amino acids from water to an organic solvent (5). Experience has shown that such calculations generally give qualitative trends that are similar to those obtained with the holistic MZ scale, but they lack the details that provide convincing models of protein functionality (7,9). So it is here: the seven nearly equal steps found with MZ become widely unequal, and sometimes even reverse sign, with the KD scale (5); it appears that the overall signal/noise ratio has dropped by at least a factor of 4 from the unifying holistic MZ scale to the fragmented reductionist KD scale, and experience with other reductionist examples suggests that their results would be equally noisy (7).

For completeness the final KD results are included in Table 1, and it might appear that they are quite similar to those with the MZ scale. However, this qualitative similarity is deceptive: it occurs because these abbreviated results are based on multiple averages. The key point of equal steps for distant surface sites for the MZ scale, but not for the KD scale, is not shown here, but it can easily be checked by the reader using hydrophobicity tables given elsewhere (7). One cannot infer Eqn. (1) from these multiply averaged results, but it does appear to be quite natural when one studies the nearly equal individual contributions to the two subsets $<<\Psi3>_M>$ and $<<\Psi3>_{M*}>$ (but only with the MZ scale, not with the KD scale).

**Disulfide Bonds in HEW and Human Lysozyme**

Compared to HEW, human lysozyme has 72 conserved sites. The 58 mutations include many innocuous ones, but even so the number of mutations is too large to be treated as perturbations, as were those of Turkey and Pheasant. Instead we focus on comparing the hydro(phobic,philic) extrema. On various reductionist hydrophobicity scales, the most hydrophobic amino acids are either Ile, Trp, or Cys (4). Because Cys is the most hydrophobic amino acid in the MZ scale, most of the MZ hydrophobic extrema are associated with disulfide C-C bonds. However, we do not list $<\Psi3(C)>$ for each C, but rather the contextual hydrophobic extremum of $<\Psi3(Y)>$, where C may be either X,Y or Z of an XYZ sequence – in other words, C can be at the center of the three-residue extremal sequence or a nearest neighbor. The advantage of concentrating on these contextual C hydrophobic extrema is brought out by comparing the eight extremals for HEW and human lysozyme (Table II). In HEW the extrema show no special properties, but in human lysozyme they separate nicely into two groups – **strong** ($<\Psi3> > 0.213$) and weak ($<\Psi3> < 0.198$) extrema. Is that all? No there's more, and it is spectacular: the two human subgroups pair off exactly, with disulfide bonds formed only between (unlike) **strong** and weak extrema ((6-**128**), (**30**-115), (**65**-81), and (77-**95**)). Thus the four disulfide bonds of human lysozyme are maximally similar (in a set-theoretic sense), whereas the four disulfide bonds of HEW exhibit no special properties. This complete separation of the eight C's of human lysozyme into two cross-bonded subgroups could provide the simplest indication of the cooperative mechanisms responsible for maximal lytic and glycol chitic activity of human lysozyme.

Comparison of these human lysozyme results obtained with the holistic MZ scale with those obtained with the reductionist KD scale brings out their significance. Again with the KD scale there are two groups of four extrema each, there is a gap between strong ($<\Psi3> > 0.201$) and weak ($<\Psi3> < 0.185$) extrema groups, and it has a similar magnitude. However, this time YW[64]C (which was in the strong group with the MZ scale) has switched to the weak group, while LS[80]C (which was in the weak group with the MZ scale) has switched to the strong group. Thus the four disulfide bonds consist of two bonds each between like and unlike, or like and like, and with the KD scale there is no qualitative difference between the disulfide bonds of HEW and human lysozyme.

Because Cys has the largest hydrophobicity in the MZ scale, disulfide bonds are ideal markers for hydrophobic lysozyme extrema. The situation is less simple for hydrophilic extrema. When one plots $<\Psi3>$ for HEW and human lysozyme (not shown here), there are long sequences of good agreement, even though the individual amino acids are often different. The two outstanding patches of hydrophilic disagreement are 42-44, where $<\Psi3>$ ~ 0.15 (human, hydroneutral) and 0.10 (HEW, hydrophilic), and 71-74, where $<\Psi3>$ ~ 0.17 (human, hydrophobic) and 0.115 (HEW, hydrophilic). Both patches are on the surface of the left lobe, and both involve hydrophobic stiffening of the human lysozyme surface patches relative to those of HEW lysozyme. This stiffening may contribute to stabilizing cooperative human lysozyme-lysozyme interactions in lytic activity (see below).

**Wild Human, Rat, Pig 1 and Rabbit Lysozymes**

Next one can analyze these four cases, using the best-studied case (human) as benchmark. Our simplest configuration coordinate is the separation of disulfide bonds into two subgroups, which then leads to four strong-weak human pairs. In rabbit, rat 1B and pig 1 lysozyme, $C^{77}YZ$ switches from weak to strong, spoiling one pair in rabbit. However, in rat 1b and pig 1, $XYC^{95}$, which is bonded to $C^{77}YZ$, also switches, from strong to weak, restoring this pair, so that rat 1b still has four strong-weak pairs. Finally, a third switch occurs in pig 1 ($C^{128}YZ$ from strong to weak), once again spoiling one pair. Thus human and rat 1b have four contextual disulfide bonds each between hydrophobically strong and weak CYs, while rabbit and pig 1 have only three such bonds. Looking at Table 1, we see (as guessed above) that the numbers of these bonds correlates well with trends in lytic activities of these four species.

As we would expect, except for this almost hidden correlation, there are long sequences where differences between human and the other three species hydrophobicities are small. Between human and rat 1b one notices two interesting differential patches; for 19-21, human $<\Psi3>$ is close to 0.15 (hydroneutral), whereas rat 1b is close to 0.20 (hydrophobic, comparable to Cys). Then for 71-74, rat 1b $<\Psi3>$ departs from human (0.17) and looks like HEW (0.11) – an evolutionary echo.

When we compare rabbit, pig 1 and human lysozymes, we again find long sequences where differences between human and the other two species hydrophobicities are small. However, the differences in the long sequence $C^{80}$-$C^{95}$ are striking. This sequence is an interlobe "necklace", wrapped around the glycol bond "throat" between the D and E sugar rings of lysozyme-$(NAG)_6$ complexes, and it is shown in blue in Fig. 1. Fig. 2 shows "necklace" hydrophobicities, and we see that human and rabbit hydrophobicities are similar, but pig 1 is significantly different, especially on the right lobe (88-94). In the first half of the sequence, pig 1 is more hydrophobic, and in the second half it is more hydrophilic. The water imbalance between the lysozyme left lobe (first half of sequence) and its right lobe (second half) is thus altered for pig 1 compared to human and rabbit. Next we look at Table 1, and we find that the glycol chitin activity of pig 1 is half that of all the other species.

Following such a spectacular result for the MZ holistic scale, one naturally asks, how successful is the KD result for hydrophobic trends in the long necklace sequence $C^{80}$-$C^{95}$? The KD trends are shown in Fig. 3. At first glance the results obtained with the reductionist KD scale look quite similar to those obtained with the holistic MZ scale, but close inspection reveals crucial differences. True, the two scales give similar trends between the three species for the second half (right lobe) of the sequence, but this second half success is cancelled by failure with the first half (left lobe). Using the reductionist scale alone, one would probably not be able to recognize the hydrophobic twisting of the rabbit sequence, which is correlated to its halved glycol chitin activity. Note that this necklace plays a secondary role (compared to the primary role of the conserved basic acid pair Asp$^{52}$ and Glu$^{35}$), but it is just such subtle hydrophobic effects that one cannot identify except with the holistic MZ scale.

**Lytic Activity**

Cationic residues are believed to be the largest factor in determining antimicrobial activity of peptide segments (18,19), which are also amphipathic (20). The cationic rich segment of lysozyme lies between $C^{95}$ and $C^{115}$ (see red segment in Fig. 1), and fragments of this segment exhibit lytic activity (21,22). The net charge (K + R - D - E) in HEW and Phea is +3, Turkey +4, Rabbit +3, (Pig 1 and Rat 1b) +4, and this increases to +5 in human lysozyme: within the seven-membered lysozyme family of Table 1, net charge of this segment correlates very well with lytic activity (except for Rabbit).

Because the cationic rich segment of lysozyme lying between $C^{95}$ and $C^{115}$ is located on the right lobe of the cleft in Fig. 1, one is tempted to guess that these two lobes themselves are globally amphipathic. The left lobe of human lysozyme (41-86) has <Ψ46> = 0.153, while the right lobe is more hydrophilic, with <Ψ84> = 0.148, so the guess seems to work. However, proteins are full of surprises: in pig 1 the two lobe hydrophobicities are nearly equal, and when one examines HEW, Rabbit and Rat 1b lysozymes, one finds a reverse relation, with left lobe values 0.145±0.001 and right lobe values 0.151. (Some might worry that these results are an artifact of the holistic MZ scale, but a similar (smaller) reversal occurs even with the reductionist KD scale; as usual, the MZ scale is much more accurate and shows a larger effect.)

It appears that the amphipathic left-right lobe reversal reflects fundamental differences in lysozyme functionality. In HEW, Rabbit and Rat 1b lysozymes, different amphipathic mechanisms are operative, which switch over to become the global lobe mechanism in human lysozyme, which thus differs substantially from HEW both in net charge and amphipathicity. The largest hydrophobicity differences occur between HEW and human lysozyme in two regions, H$_1$ and H$_2$: HEW N$^{44}$RNT$^{47}$ (<Ψ4> = 0.110) is much more hydrophilic than human N$^{44}$YNA$^{47}$ ($<\Psi 4> = 0.151$), and HEW $G^{71}SRN^{74}$ ($<\Psi 4> = 0.111$) is much more hydrophilic than human $G^{71}AVN^{74}$ ($<\Psi 4> = 0.166$). Moreover, the HEW $R^{45} \rightarrow$ human $Y^{45}$ and HEW $R^{73} \rightarrow$ human $V^{73}$ exchanges reduce the positive charges in both $H_1$ and $H_2$ of human lysozyme, thus enhancing the effectiveness of the dipolar (cationic,hydrophobic) synergistic lytic mechanism for porin (or holin) formation (3). In the other animal lysozymes the hydrophilic R positive charge is restored in $H_2$ by human $G^{71} \rightarrow$ (Rabbit, Rat 1b and Pig 1) $R^{71}$, thus producing an intermediate amphipathicity with only one R exchanged region $H_1$. It is striking and surprising that evolution did not alter the global lobe amphipathic mechanism gradually (as suggested by Table 1), nor abruptly between birds and placental animals (as one might have expected on traditional biological grounds), but instead left most of the animals with only an HEW $H_1$ local mechanism, and provided only humans with the global lobe mechanism with two localized $H_1$ and $H_2$ left lobe regions.

## Synthetic Point Mutations

Mutational experiments have been performed to support and analyze the details of the nearby basic acid pair $Asp^{52}$ and $Glu^{35}$ interactions, all with spectacular success: generally speaking, modifications of either residue are sufficient to destroy both lytic and enzymatic activity. A number of mutational studies have altered lysozyme stabilities (and sometimes even enhanced them), but it is rare for activities to be enhanced rather than destroyed (14,18). This is scarcely surprising, as the proteins have evolved to optimize their activities while maintaining merely sufficient stability. Thus the cores of lysozyme proteins are nearly perfectly conserved, while chemical trends in wild-type protein properties (previously attributed to unspecified differences in sequences (14) and internal backbone stiffness (18,19)), are caused by long range hydrophobic interactions of the types identified here with the holistic MZ scale.

The subject of folding pathways lies outside the scope of this discussion (unfolding is usually dysfunctional), but the difference between studying hydrophobic sequential profile trends in wild proteins and synthetic effects of point mutations is evident in modeling the folding kinetics of lysozyme compared to bovine α-lactalbumin (BLA). HEW folding kinetics are argued to be dominated by long range forces on the basis of sequence homologies (including hydrophobic profile homologies) with BLA, which is stabilized by short range Coulomb interactions associated with Ca binding (20). Mutations of the HEW $W^{62}W^{63}$ suggested that this pair, with its four side group rings, could play an essential role, analogous to Ca ions, as $W^{62}$ is replaced by I or L in BLA. However, $W^{62}$ is also replaced by Y (one less ring) in animal and insect lysozymes (13), already altering packing and probably not affecting folding. Moreover, unlike HEW, the hydroprofile of BLA shows much the same features with short range reductionist scales as with the long range MZ scale. This means that other factors (such as surface charge (21)) are involved in ion binding. Perhaps the unexpected (accidental?) HEW-BAL sequence homologies can be partially rationalized in terms of the long range amphipathic lysozyme interactions discussed above.

**Human Defensins**

Many antimicrobial peptides are known (22), but the best known are probably defensins, which contain the largest number of Cys (six) and disulfide bonds (three), and most closely resemble lysozyme *c*. Defensins are small (30-45 amino acids) cationic peptides found in many organisms (23). All defensins have amphiphilic properties, which are central for antimicrobial activities of the proteins. Defensins are stabilized by six Cys and three disulfide bonds, a much larger number of Cys than the two that are typical of proteins with < 100 residues (24), but in other antimicrobial peptides of similar size the number of Cys (disulfide bonds) decreases even to two (one) (22). In defensins data indicate that the main function of the disulfides may be to protect the backbone from proteolysis during biosynthesis and in protease-containing microenvironments where they function as effector molecules. Mutagenesis of disulfide bonds in α and β defensins produced analogs with *in vitro* bactericidal activities equal to or greater than that of the native peptide (22). This conclusion is consistent with the lysozyme results described above, where lytic activity is optimized by cooperativity of surface residues, while the interior residues are stabilized by the topology imposed by the disulfide bonds.

Hydroanalysis concerns itself first with distinguishing between α and β defensins, which have different disulfide bond topologies [α:1-6,2-4,3-5; β:1-5,2-4,3-6], which in turn bring the C and N terminals closer together in α than in β (22). Strongly hydrophilic cationic bonding of Arg and Lys in defensin dimers to water channels facilitates formation of 25 A transmembrane pores (24). It is puzzling that the Arg/Lys ratio in mammalian α defensins is ~ 9:1, whereas in β defensins it is 4/6 (25). On the MZ scale Lys is only slightly more hydrophilic than Arg, and the side chains have similar lengths. However, Arg contains three amide units [$(NH)_2 NH_2$], whereas Lys has only one [$NH_2$]. Each of the NH units in Arg is more flexible than the corresponding $CH_2$ units in Lys, enabling Arg side chains to compensate flexibly for the more rigid C and N terminal backbone proximity characteristic of α defensins.

Chemical trends in the structures and properties of several α neutrophil defensins (HNP3, HNP4, HD5 and HD6) are puzzling. The puzzle is that the static differences between the structures (which share three conserved disulfide bridges and one salt bridge) and presumably their stabilities are small, but there are substantial differences in antimicrobial activity. The static structural similarities persist to the solvent accessible, buried and Debye-Waller profiles of the dimerized crystal structures (Fig. 1 of (26)). However, the self-similar MZ hydrophobicity scale explores implicit (long-range) hydroelastically mediated protein-protein interactions, which are exactly what are involved in forming 25 A transmembrane pores (24). Hydrophobic plots show both similarities and differences between α neutrophil defensins; an example is given in Fig. 4. Examining these plots, one finds that the key features are the hydrophilic extrema primarily associated with charged Arg-Arg pairs, and sometimes with neutral Arg and (Asp or Glu) pairs. These hydrophilic extrema shift positions around conserved C, E or G sites, and should be the key to shifts in antimicrobial activity. (Docking of large cargoes for nuclear transport is based on hydrophilic extrema (nuclear localization sequences, NLS) dominated by Lys and Arg residues (27)).

Overall HD5 exhibits the largest antimicrobial activity of these α defensins (28): can we use this to extend our understanding of the lytic mechanism? Looking at the hydrophobic $<\Psi3>$ profile of HD5, one finds that the hydrophilic minimum occurs at sites 13-15, RES. With the other three examples, this minimum is always associated with the RR cationic charged pair, but here the RE pair is neutral. The net charge for each sequence is +2 in the other three examples, but it is +4 for HD5. Thus HD5 is the most cationic α defensin, but in addition it possesses a unique neutral NLS. Such a sequence could be quite helpful in constructing hydrophilic pair bridges between defensin dimers that then form annular porins (24), as neutral bridges are more easily formed than ones between like charges. Note that this bridging interaction is a characteristic *in vitro* property, not observable crystallographically.

Although insect defensins lie outside the scope of this paper, one can use royalisin (from royal bee jelly) to make an important point. Unlike animal defensins, which terminate near the Cys 6 residue, royalisin contains an additional 12–residue C-terminal sequence which presumably assists its lytic activity (29). The most prominent feature of this unique sequence is the strong hydrophilic minimum associated with DKR (charge only +1) near its end. It appears from this example and from the other cases studied here that cationic charge and hydrophilic nuclear localization sequences make complementary contributions to lytic activity, and that an increase in the latter can compensate a decrease in the former.

**Conclusions**

The differences between HEW and human lysozymes *c* (as well as the other five animal lysozymes discussed here) are inaccessible to most theoretical structural probes, not only MDS using CFF, but even more sophisticated "soft mode" or principal component dynamical methods designed to identify domains and hinges, as the $C_\alpha$ coordinates of these lysozymes are superposable to 0.65 A (30). As discussed in (7), there are profound differences between stability and functionality. Conserved sites, including even conserved disulfide and salt bridges, as well as 30-40% sequence conservation, yielding almost identical backbone folds, primarily assures protein stability within a given family of proteins (lysozymes or defensins), while leaving unexplained chemical trends in functionality associated with nonconserved sites.

The MZ hydrophobicity scale includes the effects of self-similarity and self-organized criticality, and this enables it to explain chemical trends in functionality that are inaccessible to most theoretical structural probes, as well as less accurate scales that describe protein stability and transition states associated with dysfunctional protein unfolding. However, the successes described here and elsewhere leave an important question unanswered: just why are proteins so close to self-organized criticality (SOC)? One can conjecture that power-law evolution of solvent-accessible surface areas describes compacted yet still stress-free protein networks (7,9,31). This mathematical evolution has been made possible because the richness of the amino acid menu has made it possible for proteins to adapt their (several) functionalities optimally, and specifically through water-mediated interactions. This picture not only has attractive fundamental aspects as regards Leventhal's paradox (7), but it also leads directly to tangible results, in contrast to energy landscape formalisms based on classical force fields (31).

**Methods**

The MZ and KD hydrophobicity scales are tabulated elsewhere (7,11). KD has been rescaled so that its range and hydroneutral midpoint matches those of the MZ scale, which facilitates comparisons of the effectiveness of the KD scale with that of the MZ scale based on SOC.

| Lysozyme | Lytic act. | Glycol chitin act. | Sugar bind. | $\langle\langle\Psi3\rangle_M\rangle$ | $\langle\langle\Psi3\rangle_{M*}\rangle$ | $\langle\langle\Psi3\rangle_M\rangle$ | $\langle\langle\Psi3\rangle_{M*}\rangle$ |
|---|---|---|---|---|---|---|---|
| HEW | 100 | 100 | 71400 | 0.143 | 0.143 | 0.148 | 0.133 |
| Pheas. | 123 | 82 | 55600 | 0.168 | 0.154 | 0.169 | 0.138 |
| Turkey | 176 | 80 | 8300 | 0.159 | | 0.161 | |
| Rabbit | 204 | 99 | 17500 | | | | |
| Pig I | 245 | 45 | 8300 | | | | |
| Rat 1b | 255 | 99 | 9200 | | | | |
| Human | 396 | 110 | 10000 | | | | |

Table 1. The entries for lytic acivity (against gram-positive bacteria) and activity against glycol chitin (a soluble linear homopolysaccharide) are normalized to 100% for HEW (from (14)). The sugar binding ability is represented by the association constant $K_A$ against (NAG)$_3$. The three-site average hydrophobicities are calculated with the "exact" MZ scale for two subsets (M and M*) of Pheas. and Turkey mutations relative to HEW (see text).

| Peak | <Ψ3> MZ | MZ C Strength | <Ψ3>KD | KD C Strength |
|------|---------|---------------|--------|---------------|
| MC30L | 221.3 | Strong | 218.8 | Strong |
| YW64C | 214 | Strong | 167.0 | Weak |
| VA94C | 213.7 | Strong | 220.8 | Strong |
| CG129V | 213.3 | Strong | 206.4 | Strong |
| CE7L | 179 | Weak | 183.4 | Weak |
| CH78L | 198.3 | Weak | 185.4 | Weak |
| LS80C | 181 | Weak | 201.0 | Strong |
| CQ117N | 154.7 | Weak | 135.4 | Weak |

Table 2.  Hydroanalysis of contextual disulfide (Cys-Cys) bonds in human lysozyme using the holistic MZ and reductionist KD scales.  The first four C's are bonded to the last four.

# Figure Captions

Fig. 1. A sketch of lysozyme *c*, adapted from (14). Six sugar rings (A-F) are indicated in black in the cleft between two lobes. The basic mechanism for enzyme activity is still charge exchange between conserved Glu 35 and Asp 52, but most of the other features (such as the emphasis placed on Trp) inherited from (13,14) no longer appear to be significant (15). The new points are explicit identification of the factors responsible for the species-dependent chemical trends shown in Table 1. These include the hydrophobically skewed necklace 80-90 in blue, the cationic-rich right lobe segment between C94 and C115 in red, and the two hydrophobically adjusted left lobe ears $H_1 = 44-47$ and $H_2 = 71-74$ (in green). Most of the trends in animal lytic activity are associated with amphipathic changes in net charge of these two green ears and of the cationic-rich red segment.

Fig. 2. MZ hydropatterns between C80 and C94 (the blue interlobe necklace in Fig. 1).

Fig. 3. KD hydropatterns between C80 and C94 (the blue interlobe necklace in Fig. 1).

Fig. 4. Hydropatterns for two α defensins. The deepest hydrophilic minimum for HNP3 is $E^{14}R^{15}R^{16}$, while that for HNP4 is $C^{10}R^{11}R^{12}$. Note that the former has a net charge of only +1 and is adjacent to a salt bridge, while the latter is adjacent to a disulfide bond, which creates large differences in antimicrobial activities.

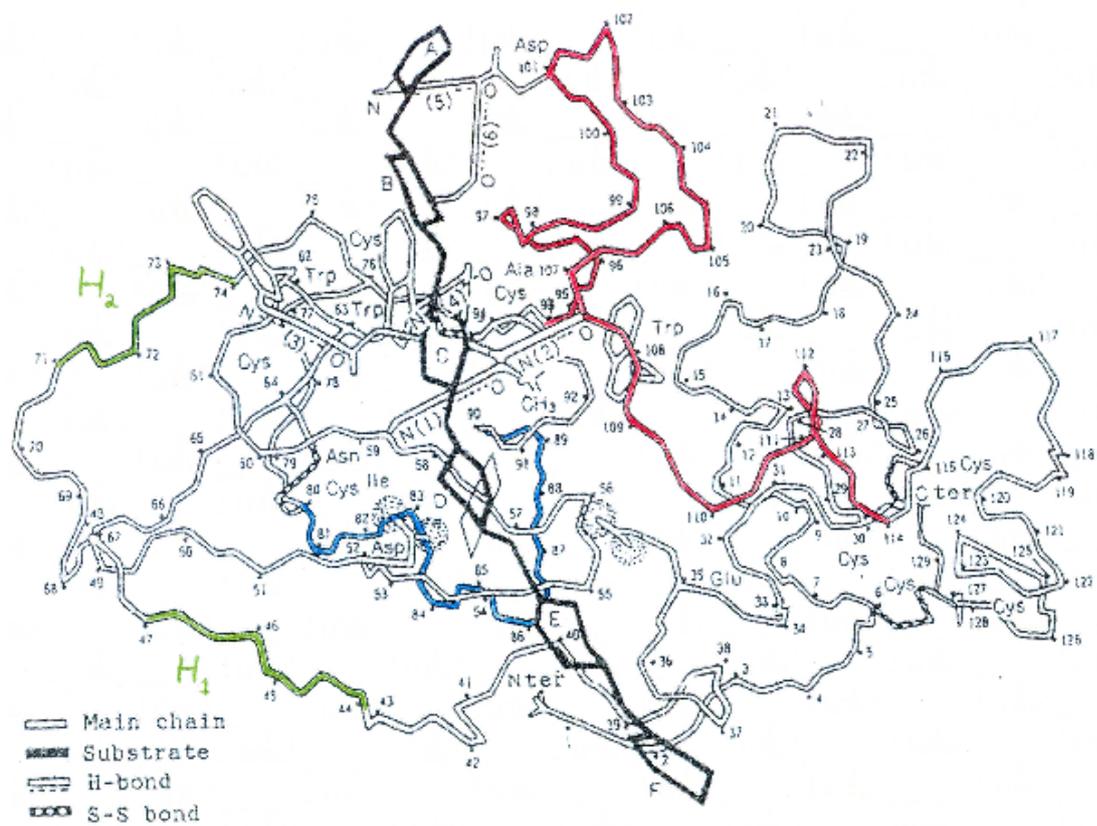

Fig. 1.

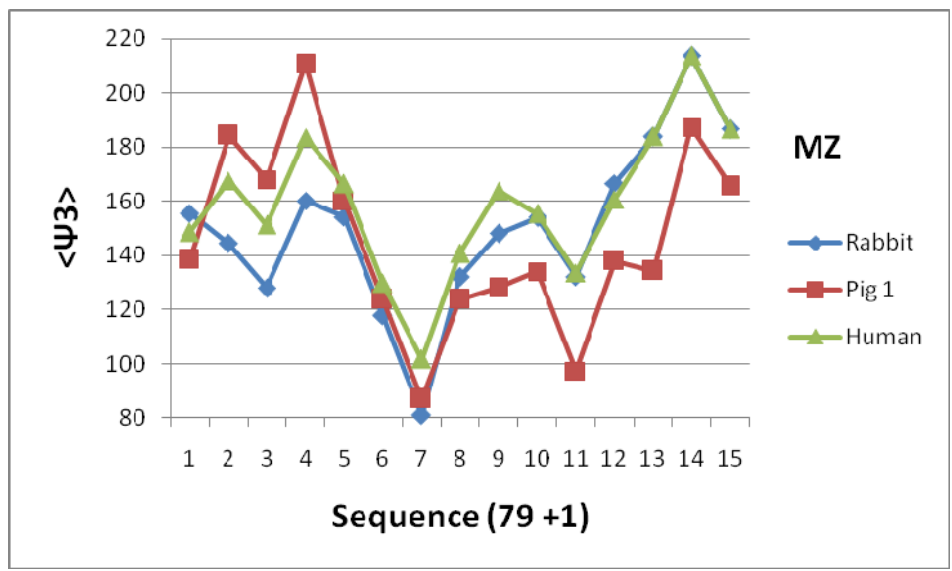

Fig. 2.

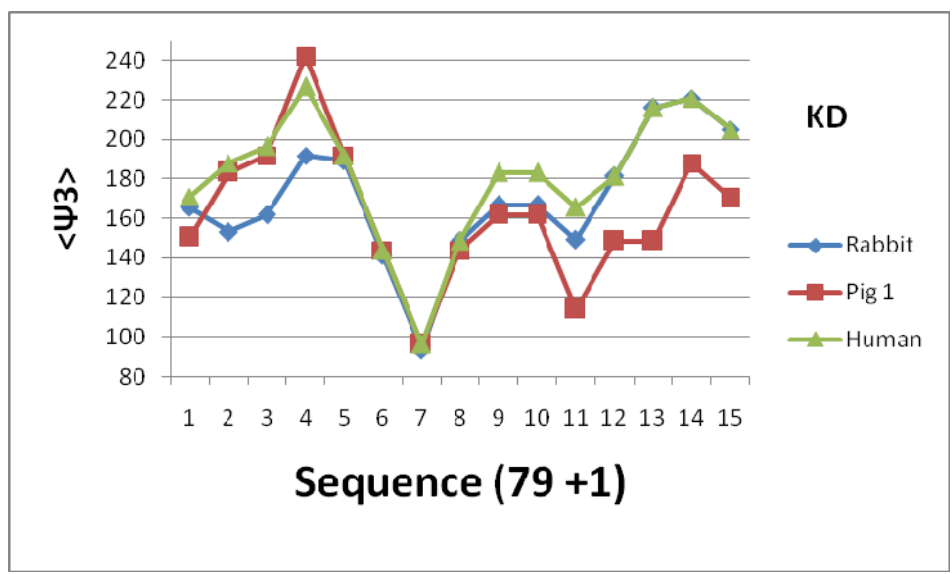

Fig. 3.

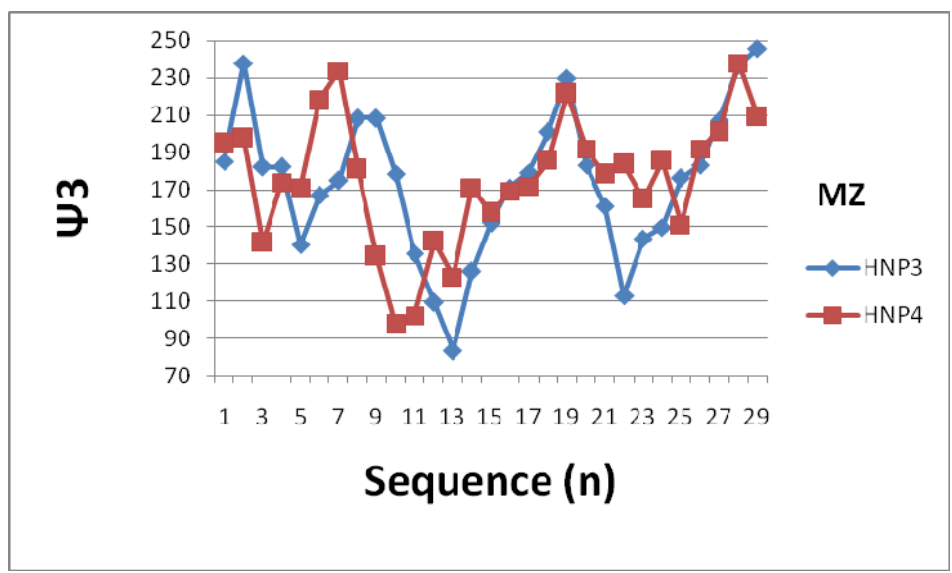

Fig. 4.